# THE HYDROGEN ATOM ACCORDING TO WAVE MECHANICS IN CARTESIAN COORDINATES

*J. F. Ogilvie*∗

Centre for Experimental and Constructive Mathematics, Department of Mathematics,
Simon Fraser University, Burnaby, British Columbia V5A 1S6 Canada

Escuela de Química, Universidad de Costa Rica, Ciudad Universitaria Rodrigo Facio,
San Pedro de Montes de Oca, San José, 11501-2050 Costa Rica

**Abstract**

A partial separation of the variables is practicable for the solution of Schroedinger's temporally independent equation in cartesian coordinates $x,y,z$, which yields moderately simple algebraic formulae for the amplitude functions involving quantum numbers $k$, $l$, $m$, the same as in spherical polar coordinates. The properties of angular momentum are thus achieved with no angular variable. Several plots of surfaces of constant $\psi(x,y,z)$ are presented to illustrate the resemblance of the shapes of these surfaces to the shapes of surfaces of $\psi(r,\theta,\phi)$ with the corresponding quantum numbers.

## I  INTRODUCTION

In 1926 Schroedinger initiated wave mechanics, which is now recognised to be one method among many that collectively constitute quantum mechanics, with a solution of his equation for the hydrogen atom in spherical polar coordinates [1]. For physicists this achievement became a second or third such method that might be applied to treat various problems more or less related to sundry experiments in quantum physics, but for chemists this solution provided, rightly or wrongly, a basis for a conceptual approach that pervades much chemical description of the structure and reactions of molecules and materials in terms of *orbitals*. Although the latter are formally merely mathematical entities -- algebraic formulae, in the minds of many chemists whose understanding of the mathematical basis is weak, they have become imbued with properties and significance far transcending their algebraic reality [2]. An orbital is logically defined as an amplitude function (Schroedinger's term) that results from a solution of Schroedinger's temporally independent equation for an atom with one electron. That equation is solvable with a complete separation of the variables in spatial coordinates in four systems and momentum coordinates in six systems [3]; those solutions in spherical polar, paraboloidal, ellipsoidal and spheroconical spatial coordinates have been presented in detail, with many plots of surfaces of constant amplitude to reveal the geometric properties, in preceding papers in this journal [4 - 8]. Of the six systems feasible for a complete solution in momentum space [3], only two systems have been treated other than a formal recognition of their existence; in one case the coordinates obtained from Fourier transformation from spherical polar spatial coordinates were not orthogonal [9], which would complicate calculations based on these functions, but, when orthogonality was imposed [10], the resulting formulae include a Dirac delta function for one coordinate; the latter property is questionable for a quantity of which its square must possess a physical meaning [11]. In the second case, as a transformation from paraboloidal coordinates [12], the resulting toroidal coordinates, equivalent to cylindrical coordinates [3], likewise lack that orthogonality property.

Distinct from a complete separation of variables in a coordinate space, a partial separation is practicable in cartesian coordinates -- $x$, $y$, $z$ according to common convention; although the results deserve to be known for applications in both physics and chemistry, the communication [13] of this worthy calculation received no citation in the literature for more than a half century, until our recent recollection of its existence [5]. Here we recall and extend the derivation of the resulting algebraic functions and present some plots of their surfaces of constant amplitude to demonstrate the utility of that derivation.

## II   DERIVATION OF THE GENERAL AMPLITUDE FUNCTION

Fowles began his derivation [13] with an assumption that amplitude function $\psi(x,y,z)$ had the form of a function of four variables, $x$, $y$, $z$ and $r = \sqrt{x^2 + y^2 + z^2}$, so $f(x,y,z,r)$, and further that the part of $f$ dependent on $r$ is separable, so $f(x,y,z,r) = F(x,y,z) R(r)$; this device enables a partial separation of the variables. Under these conditions and after some algebraic manipulation, Schroedinger's equation becomes written (with mathematical software *Maple*) in these coordinates as

$$R(r) \left( \left( \frac{\partial^2}{\partial x^2} F(x, y, z) \right) + \left( \frac{\partial^2}{\partial y^2} F(x, y, z) \right) + \left( \frac{\partial^2}{\partial z^2} F(x, y, z) \right) \right)$$

$$+ \frac{2 \left( \frac{d}{dr} R(r) \right) \left( x \left( \frac{\partial}{\partial x} F(x, y, z) \right) + y \left( \frac{\partial}{\partial x} F(x, y, z) \right) + z \left( \frac{\partial}{\partial x} F(x, y, z) \right) \right)}{r}$$

$$+ F(x, y, z) \left( \frac{d^2}{dr^2} R(r) \right) + \frac{2 F(x, y, z) \left( \frac{d}{dr} R(r) \right)}{r} + \frac{8 \pi^2 \mu (E - V(r)) R(r)}{h^2} = 0 \quad ,$$

in which the potential energy $V(r)$ that takes into account the coulombic attraction appears in the ultimate term on the left side of the equality. Fowles proceeded to seek solutions such that $F(x,y,z)$ satisfies the Laplace equation, $\nabla^2 F_l(x,y,z) = 0$; this assumption might be justified if one consider that radial function $R(r)$ takes into account the coulombic term in the customary hamiltonian operator, leaving the laplacian operator involving the cartesian coordinates. Differentiation of $F_l(x,y,z) = (a\,x + b\,y + c\,z)^l$ provides a proof that this formula is a solution provided that $a^2 + b^2 + c^2 = 0$; furthermore, for $F_l(x,y,z)$ to be singly valued, $l$ must be integer.  Under these conditions Schroedinger's equation reduces to this radial equation.

$$\left( \frac{d^2}{dr^2} R(r) \right) + \frac{2\,l + 1}{r} \frac{d}{dr} R(r) + \frac{8 \pi^2 \mu (E - V(r)) R(r)}{h^2} = 0$$

With coulombic potential energy in SI units for an atom of atomic number Z having only one electron,

$$V(r) = -\frac{Z e^2}{4\pi \varepsilon_0 r},$$

on introducing dimensionless variable $\rho = \alpha r$ and eigenvalue parameter $\lambda = \dfrac{Z e^2 \sqrt{\mu}}{2 \varepsilon_0 h \sqrt{-2E}}$, in which $\alpha = \dfrac{8\pi^2 \mu Z e^2}{h^2 \varepsilon_0 \lambda}$, and on letting $R(r) = e^{\left(-\frac{\rho}{2}\right)} L(\rho)$, the radial equation becomes transformed into

$$\rho \left(\frac{d^2}{d\rho^2} L(\rho)\right) + (2l + 1 - \rho)\left(\frac{d}{d\rho} L(\rho)\right) + (\lambda - l - 1) L(\rho) = 0$$

This differential equation has well behaved solutions in which appear associated Laguerre polynomials $L_k^{(2l+1)}(\rho)$ with $\lambda = n = k + l + 1$, a positive integer. Eigenvalues $E_{k,l}$ hence conform to the conventional formula whereby the energy of a bound state is proportional to the inverse square of a positive integer.

$$E_{k,l} = -\frac{\mu Z^2 e^4}{8 h^2 \varepsilon_0^2 (k+l+1)^2}$$

The corresponding eigenfunctions are derived explicitly, as expressed directly with mathematical software *Maple* to ensure an accurate presentation [14], in this form that is slightly modified from the solution that Fowles published to take into account the SI convention; $a_\mu$ denotes the effective Bohr radius $a_\mu = \dfrac{h^2 \lambda}{8 \pi^2 Z \mu e^2}$ with $a_\mu = \dfrac{m_e a_0}{\mu}$ in terms of reduced mass $\mu$ for the electronic and nuclear masses.

$$\psi_{k,l}(x,y,z) := \sqrt{\frac{(k+2l+2)!(k+2l+2)}{a_\mu k!}} (ax + by + cz)^l e^{\left(-\frac{\sqrt{x^2+y^2+z^2}}{a_\mu(k+l+1)}\right)}$$
$$\text{LaguerreL}\left(k, 2l+1, \frac{\sqrt{x^2+y^2+z^2}}{a_\mu(k+l+1)}\right) \bigg/ ((-1)^{(2l+1)} (k+l+1)! (2+2l)!)$$

This general formula, which includes a normalising factor derived from only the radial part containing the exponential function and the associated Laguerre polynomial, contains quantum numbers $k,l$, that arise from the solution for radial function $R(r)$. A requisite of a solution of Schroedinger's equation for a bound system is that the amplitude function must decay to zero as $r \to \infty$; this condition is satisfied if the Laguerre function becomes a polynomial of finite order, which is fulfilled when parameters $k$ and $l$ take discrete values of non-negative integers. To define a third quantum number according to the applicable criteria [3], we consider the angular momentum of states of the hydrogen atom.

## III  ANGULAR MOMENTUM AND AMPLITUDE FUNCTION

The two most important properties to characterise a state of any atom are its energy and its angular momentum. For only an atom with one electron, such as hydrogen, is the energy of a bound state synonymous with an energy quantum number. Having obtained the energy explicitly defined in an expression above, we consider the angular momentum, a vectorial quantity $M$. For a component of angular momentum about axis $z$, we have, with $i = \sqrt{-1}$ and Planck constant $h$,

$$M_z \psi = -\frac{i h}{2 \pi} \left( x \left( \frac{\partial}{\partial y} \psi \right) - y \left( \frac{\partial}{\partial x} \psi \right) \right)$$

with analogous formulae for $M_x \psi$ and $M_y \psi$ in which the axes are permuted cyclically. Taking into account that factor $R(r)$ commutes with the angular momentum operator, we obtain, using the properties of the derivatives applied in the generation of the amplitude functions above,

$$M_z F(x, y, z) = -\frac{i h}{2 \pi} \left( x \left( \frac{\partial}{\partial y} F(x, y, z) \right) - y \left( \frac{\partial}{\partial x} F(x, y, z) \right) \right)$$

As $R(r)$ commutes also with operator $M^2 = M_x^2 + M_y^2 + M_z^2$, we obtain, equating $F(x,y,z)$ with $F_l(x,y,z)$ that is a solution of the Laplace equation from above,

$$M^2 F_l(x, y, z) = -\left(\frac{h}{2\pi}\right)^2 \left( \left( x \left( \frac{\partial}{\partial y} F_l(x, y, z) \right) - y \left( \frac{\partial}{\partial x} F_l(x, y, z) \right) \right)^2 \right.$$
$$\left. + \left( y \left( \frac{\partial}{\partial z} F_l(x, y, z) \right) - z \left( \frac{\partial}{\partial y} F_l(x, y, z) \right) \right)^2 + \left( z \left( \frac{\partial}{\partial x} F_l(x, y, z) \right) - x \left( \frac{\partial}{\partial z} F_l(x, y, z) \right) \right)^2 \right)$$

which, on substituting the formula for $F_l(x,y,z)$, yields the result

$$M^2 F_l(x, y, z) = \left(\frac{h}{2 \pi}\right)^2 l (l + 1) F_l(x, y, z)$$

This formula indicates that the eigenvalues of $M^2$ are $\left(\frac{h}{2\pi}\right)^2 l(l+1)$, with no preferred axis. For a component of $M$ such as $M_z$, implementing the derivatives [13, 14] yields

$$M_z F_l(x, y, z) = -\frac{i h l (-a y + b x) F_l(x, y, z)}{2 \pi (a x + b y + c z)}$$

For $F_l(x,y,z)$ to satisfy the Laplace equation according to the condition that $a^2 + b^2 + c^2 = 0$ implies that $a,b,c$ must be defined in terms of two arbitrary complex numbers $u$ and $v$, as follows:

$$a = u^2 - v^2, \quad b = -i (u^2 + v^2), \quad c = -2 u v$$

$F_l(x,y,z)$ becomes

$$F_l(x, y, z) = (u^2 (x - i y) - v^2 (x + i y) - 2 u v z)^l$$

Expressed in this manner, $F_l(x,y,z)$ as a homogeneous polynomial of degree $2l$ in $u, v$ contains $2l + 1$ terms, the coefficients of which are polynomials of degree $l$ in $x, y, z$.

$$F_l(x, y, z) = \sum_{m=-l}^{l} u^{(l-m)} v^{(l+m)} Q_{l,m}(x, y, z)$$

In table 1 we present the values of $Q_{l,m}(x,y,z)$ for $l$ up to 4 and all feasible values of $m$. Because $u$ and $v$ are arbitrary complex numbers, each $Q_{l,m}$ is a solution of the Laplace equation and, according to the assumptions above, is hence a suitable eigenfunction for the solution of the hydrogen atom in wave mechanics in cartesian coordinates. In general,

$$-i \left( x \left( \frac{\partial}{\partial y} Q_{l,m} \right) - y \left( \frac{\partial}{\partial x} Q_{l,m} \right) \right) = m\, Q_{l,m}$$

in which $m$ takes values of integers from $-l$ to $l$. In this representation in terms of $Q_{l,m}$ according to the choice of conditions above, component $z$ of angular momentum, i.e. $M_z$, has eigenvalue $\frac{m h}{2 \pi}$, which sets the third quantum number. The total amplitude function, incompletely normalised, in terms of cartesian coordinates and containing three quantum numbers $k,l,m$ hence becomes

$$\psi_{k,l,m}(x, y, z) = \sqrt{\frac{(k + 2l + 2)!\,(k + 2l + 2)}{k!\, a_\mu}} \, Q_{l,m}(x, y, z)\, e^{\left( -\frac{\sqrt{x^2 + y^2 + z^2}}{a_\mu (k + l + 1)} \right)}$$

$$\text{LaguerreL}\left( k, 2l + 1, \frac{\sqrt{x^2 + y^2 + z^2}}{a_\mu (k + l + 1)} \right) \Big/ ((-1)^l (2 + 2l)!\,(k + l + 1)!)$$

### IV  PLOTS OF AMPLITUDE FUNCTIONS

We present plots of selected amplitude functions in cartesian coordinates to display the geometric properties of surfaces of constant $\psi$ at a value chosen to be about 1/100 of the maximum value of $\psi$ for that particular function; such a criterion implies that about 0.995 of the total electronic charge density is found within a corresponding surface of $\psi^2$. The scale of each axis has unit $a_\mu$, the Bohr radius adjusted for the reduced mass, approximately $5.2 \times 10^{-11}$ m. Figure 1 shows first the surface of amplitude function $\psi_{0,0,0}(x,y,z)$, which has a spherical shape and diameter identical with those of the corresponding surfaces of functions for the ground state in all four coordinate systems in which the spatial variables are completely separable [5 - 8].

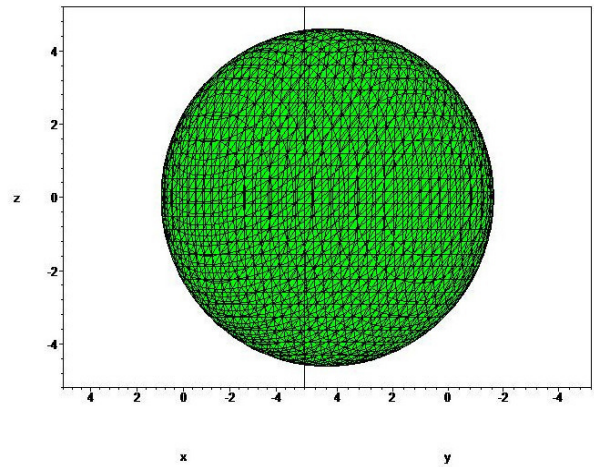

Figure 1  Surface of amplitude function $\psi_{0,0,0}(x,y,z) = \sqrt{\dfrac{1}{a_\mu}}\, e^{\left(-\dfrac{\sqrt{x^2+y^2+z^2}}{a_\mu}\right)}$ at a constant value of ψ taken to be 1/100 of the maximum amplitude.  Here, and in succeeding plots in three pseudo-dimensions, the unit of length along each coordinate axis is Bohr radius $a_\mu$.

The next surface, in figure 2, is that of amplitude function $\psi_{0,1,0}(x,y,z)$, which consists of two roughly hemispherical lobes with rounded edges, one with positive phase and the other with negative phase. Amplitude functions $\psi_{0,1,1}(x,y,z)$ and $\psi_{0,1,-1}(x,y,z)$ are complex, requiring separate plots of their real and imaginary parts; those surfaces of the real parts have shapes and sizes identical with those of the surface in figure 2 but are cylindrically symmetric about axes *y* and *x* respectively, instead of axis *z* for $\psi_{0,1,0}(x,y,z)$, whereas the respective imaginary parts are symmetric about axes *x* and *y*.

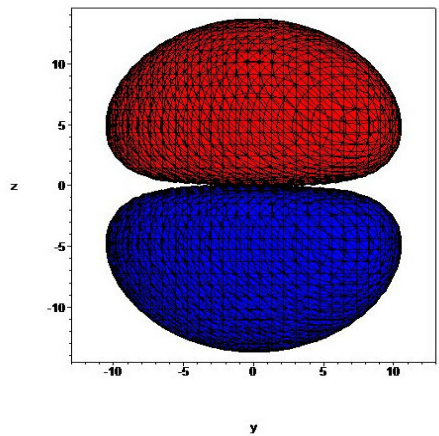

Figure 2 Surface of amplitude function $\psi_{0,1,0}(x, y, z) = \dfrac{\sqrt{\dfrac{6}{a_\mu}}\, z\, e^{\left(-\dfrac{\sqrt{x^2+y^2+z^2}}{2 a_\mu}\right)}}{6}$

In figure 3 the surface of real amplitude function $\psi_{0,2,0}(x,y,z)$ consists of three lobes, one being a circular torus of symmetric cross section, having its centre at the origin, of negative phase that separates two conical spheroidal lobes of positive phase. Of four other functions with $k=0$ and $l=2$, of which $m = -2, -1, 1, 2$, each function has both real and imaginary parts; each surface comprises four lobes of slightly conical spheroidal shape that lie along or between the cartesian axes and with the apices of their cones directed toward the origin.

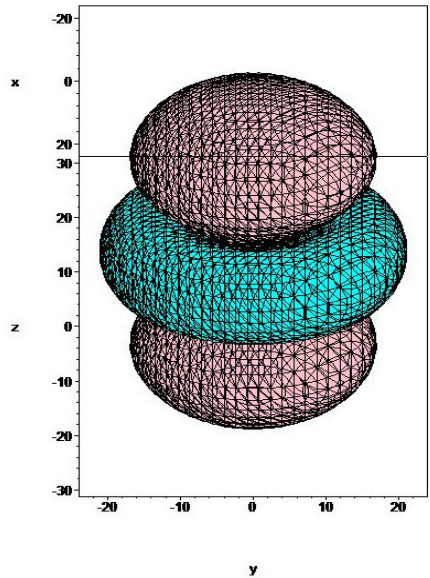

Figure 3 Surface of amplitude function

$\psi_{0,2,0}(x, y, z) = -\dfrac{\sqrt{\dfrac{30}{a_\mu}}\,(x^2+y^2-2z^2)\, e^{\left(-\dfrac{\sqrt{x^2+y^2+z^2}}{3 a_\mu}\right)}}{180}$ , comprising a circular torus that separates two conical spheroidal lobes

In figure 4 we show the surface of real amplitude function $\psi_{0,3,0}(x,y,z)$ that consists of four lobes; two of these are circular tori of opposite phases and unsymmetric cross section but surrounding axis $z$. and the other two are conical spheroids, also of opposite phases, with the apices of their cones pointing toward the origin.

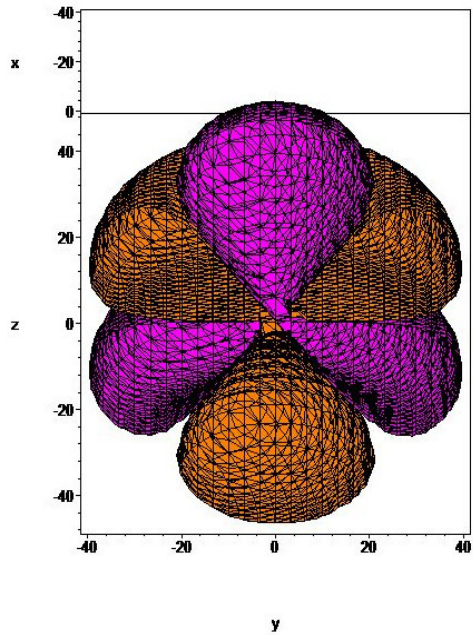

Figure 4  Surface of amplitude function
$$\psi_{0,3,0}(x, y, z) = -\frac{\sqrt{\frac{35}{a_\mu}}\, z\left(x^2 + y^2 - \frac{2 z^2}{3}\right) e^{\left(-\frac{\sqrt{x^2 + y^2 + z^2}}{4 a_\mu}\right)}}{840}$$
, cut open to reveal the internal structure

In figure 5 the surface of amplitude function $\psi_{0,4,0}(x,y,z)$ exhibits five lobes of which three are circular tori about axis $z$; the central torus has positive phase, like the two conical spheroidal lobes separated by the three tori, and the other two tori have negative phase.

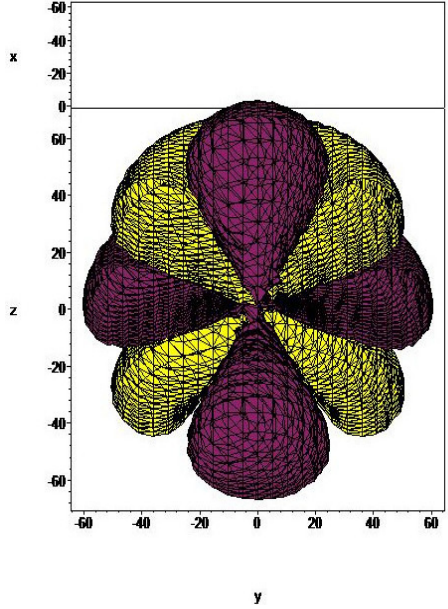

Figure 5 Surface of amplitude function

$$\psi_{0,4,0}(x,y,z) = \frac{\sqrt{\frac{70}{a_\mu}}\left(x^4 + (2y^2 - 8z^2)x^2 + y^4 - 8y^2 z^2 + \frac{8 z^4}{3}\right) e^{\left(-\frac{\sqrt{x^2+y^2+z^2}}{5 a_\mu}\right)}}{100800}$$

, cut open to reveal the internal structure

In figure 6 the surface of amplitude function $\psi_{1,0,0}(x,y,z)$ exhibits an inner sphere of positive phase surrounded by a spherical shell of negative phase. Further surfaces with radial quantum number $k > 0$ and $l > 0$ likewise possess inner lobes, but the shape of the outermost surface resembles that of the respective surface for $\psi_{0,l,m}(x,y,z)$.

## V  DISCUSSION

The derivation above, adapted and extended from work of Fowles [13], and the plots in figures 1 - 6 of the amplitude functions explicitly specified in algebraic form demonstrate that a partial separation of variables as cartesian coordinates is effective to yield algebraic formulae that possess the properties appropriate to a solution of Schroedinger's equation for the hydrogen atom, beyond the four coordinate systems in which a complete separation of variables is practicable. The plots of surfaces of the selected amplitude functions have exactly the same size and shape as their counterparts in spherical polar coordinates [5]. This property is predictable because the sets of quantum numbers -- $k,l,m$ -- are the same in both cases and because one can simply transform the coordinates directly between spherical polar and cartesian; for instance, $r \cos(\theta)$ that occurs as the angular part of $\psi_{0,1,0}(r,\theta,\phi)$ is equivalent to $z$ that occurs in $\psi_{0,1,0}(x,y,z)$. One might hence regard these amplitude functions in cartesian coordinates as variants of the corresponding amplitude functions in spherical polar coordinates, although they are derived independently. The significance of the work of Fowles [13] is that it proves that orbitals with the same geometric properties are derivable in distinct systems of coordinates. As one is more familiar with the cartesian system of coordinates than the spherical polar system, the formulae involving polynomials in cartesian components, as listed in table 1, are likely more meaningful than the products of trigonometric functions with arguments θ and φ that occur in spherical polar coordinates. Although there be no factor $\mathbf{e}^{im\phi}$ in the amplitude functions in cartesian coordinates that bestows a complex character on amplitude functions in spherical polar, paraboloidal and ellipsoidal coordinates, the imposition that $F_l(x,y,z)$ must satisfy the Laplace equation with a condition $a^2 + b^2 + c^2 = 0$ requires generally complex quantities for $Q_{l,m}(x,y,z)$; most amplitude functions $\psi_{k,l,m}(x,y,z)$ have hence both real and imaginary parts.

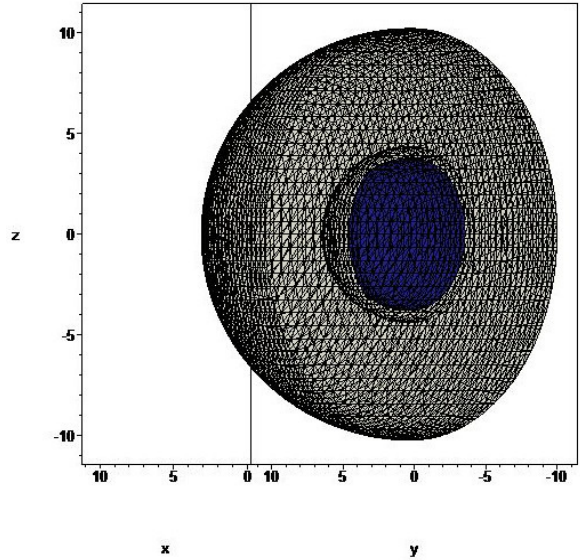

Figure 6 Surface of amplitude function $\psi_{0,1,0}(x,y,z) = \sqrt{\dfrac{1}{6\,a_\mu}}\, z\, e^{\left(-\dfrac{\sqrt{x^2+y^2+z^2}}{2\,a_\mu}\right)}$, cut open to reveal the internal structure

Whereas partial separations of coordinates are practicable for the solution of Schroedinger's equation in other systems, as one can readily test with advanced mathematical software (such as *Maple*) typically to yield one angular coordinate equivalent to φ in spherical polar coordinates, such an implementation leaves the other two coordinates in an intractable mixture; the procedure is hence unproductive of practical amplitude functions with separable variables. A claimed complete separation to solve "the Coulomb problem in cylindrical polar coordinates" is erroneous [15]. In combination with the preceding collection of amplitude functions for the hydrogen atom in four systems of coordinates with complete separation of variables [4-8], the additional amplitude functions in cartesian coordinates here hence represent the full current knowledge about these algebraic solutions for the hydrogen atom in non-relativistic wave mechanics. The full details of the derivation of these cartesian amplitude functions are available elsewhere [14].

**Table 1** Values of $Q_{l,m}(x,y,z)$

$Q_{0,0}(x, y, z) = 1$

$Q_{1,-1}(x, y, z) = x - i y$

$Q_{1,0}(x, y, z) = -2 z$

$Q_{1,1}(x, y, z) = -i y - x$

$Q_{2,-2}(x, y, z) = (i y - x)^2$

$Q_{2,-1}(x, y, z) = 4 (i y - x) z$

$Q_{2,0}(x, y, z) = -2 x^2 - 2 y^2 + 4 z^2$

$Q_{2,1}(x, y, z) = 4 (x + i y) z$

$Q_{2,2}(x, y, z) = (x + i y)^2$

$Q_{3,-3}(x, y, z) = -(i y - x)^3$

$Q_{3,-2}(x, y, z) = -6 z (i y - x)^2$

$Q_{3,-1}(x, y, z) = 3 (x^2 + y^2 - 4 z^2) (i y - x)$

$Q_{3,0}(x, y, z) = 4 z (3 x^2 + 3 y^2 - 2 z^2)$

$Q_{3,1}(x, y, z) = 3 (x^2 + y^2 - 4 z^2) (x + i y)$

$Q_{3,2}(x, y, z) = -6 z (x + i y)^2$

$Q_{3,3}(x, y, z) = -(x + i y)^3$

$Q_{4,-4}(x, y, z) = (i y - x)^4$

$Q_{4,-3}(x, y, z) = 8 z (i y - x)^3$

$Q_{4,-2}(x, y, z) = -4 (x^2 + y^2 - 6 z^2) (i y - x)^2$

$Q_{4,-1}(x, y, z) = -8 (3 x^2 + 3 y^2 - 4 z^2) (i y - x) z$

$Q_{4,0}(x, y, z) = 6 x^4 + 12 x^2 y^2 - 48 x^2 z^2 + 6 y^4 - 48 y^2 z^2 + 16 z^4$

$Q_{4,1}(x, y, z) = -8 (3 x^2 + 3 y^2 - 4 z^2) (x + i y) z$

$Q_{4,2}(x, y, z) = -4 (x^2 + y^2 - 6 z^2) (x + i y)^2$

$Q_{4,3}(x, y, z) = 8 z (x + i y)^3$

$Q_{4,4}(x, y, z) = (x + i y)^4$